\documentclass[12pt]{article}
\usepackage{pdproc}
\usepackage{amssymb}

\begin{document}

\renewcommand{\thefootnote}{\alph{footnote}}                                                                                   
         
\title{VARYING CONSTANTS\footnote{Invited talk at
the 10th international workshop on ``Neutrino Telescopes'', 11-14 March 2003 (Venice, Italy).}}
\author{Thibault DAMOUR}  
\address{Institut des Hautes Etudes Scientifiques, 35, route de Chartres, \\ 91440 
Bures-sur-Yvette, France}

\abstract{
We review some string-inspired theoretical models which
incorporate a correlated spacetime variation of coupling constants while remaining naturally
compatible both with phenomenological constraints coming from geochemical data (Oklo; 
Rhenium decay) and with present equivalence principle tests. Barring unnatural
fine-tunings of parameters, a variation of the fine-structure constant as large as that
recently ``observed'' by Webb et al. in quasar absorption spectra appears to be 
incompatible with these phenomenological constraints. Independently of any model, it 
is emphasized that the best experimental probe of varying constants are high-precision 
tests of the universality of free fall, such as MICROSCOPE and STEP. Recent claims 
by Bekenstein that fine-structure-constant variability does not imply
detectable violations of the equivalence principle are shown to be untenable.
}

\section{Introduction}

Einstein's theory of General Relativity (1915) has deeply transformed one aspect of the
general framework of physics. Before 1915, both the structure of spacetime and the laws of local 
matter interactions were supposed to be ``rigid'', i.e. given once for all, as absolute 
structures, independently of the material content of the world. Einstein's theory
introduced the idea that the structure of spacetime might be
``soft'', i.e. influenced by its material content. However, he postulated (``principle 
of equivalence'') that the laws of local physics, and notably the values of all the 
(dimensionless) coupling constants ($e^2 / \hbar c$, $m_e / m_p , \ldots$), must be 
kept ``rigidly fixed''. General Relativity thereby introduces a 
dissymetry between a ``soft'' spacetime and a ``rigid'' matter. By contrast,
one can view String theory  as a framework treating symmetrically spacetime and matter
interactions and suggesting that both of them are ``soft''. Indeed, one of the
hints of String theory is that the coupling ``constants'' appearing in the 
low-energy Lagrangian are
determined by the vacuum expectation values (VEV) of some a priori massless scalar
fields: dilaton and moduli. For instance, the VEV of the dilaton $\phi$ determines 
the basic string
coupling constant $g_s = e^{\phi / 2}$~\cite{W84}.
It is amusing to note that
this generalized correlated ``softening'' of structures (which were traditionally 
considered as independent and rigid) serendipitously shows up even in the mathematical 
notation used to represent them, through a multiple use of the letter ``$g$'': at the 
tree-level of string theory there is a link not only between {\bf g}eometry and {\bf 
g}ravitation (through the unified geometrical field ${\bf g}_{\mu\nu} (x)$), but also 
between gravitation $({\bf g}_{\mu\nu} (x))$, string coupling $({\bf g}_s (x))$, gauge 
couplings $({\bf g} (x))$ and gravitational coupling $({\bf G} (x))$. We refer here to 
a low-energy Lagrangian density of the form ($\alpha' = 1$)
\begin{equation}
\label{eqn1}
{\cal L} = \sqrt{\widetilde g} \, e^{-\phi} \left[\widetilde R + 2 \, \widetilde{\Box} 
\, \phi - (\widetilde \nabla \, \phi)^2 - \frac{1}{4} \, \widetilde F^2 + \cdots 
\right] \, .
\end{equation}
Actually such a tree-level Lagrangian (with a massless dilaton $\phi$) is in conflict 
with experimental tests of the equivalence principle. Indeed, the dilaton has 
gravitational-strength couplings to matter which violate the equivalence 
principle~\cite{TV88,DP94}. For instance, using the results of Ref.~\cite{DP94}, one derives 
that the Lagrangian (\ref{eqn1}) predicts a violation of the universality of free fall 
at the level $\Delta a / a \sim 10^{-5}$ (to be compared with the present limits $\sim 
10^{-12}$~\cite{Su94}), and a time variation of the fine-structure constant $e^2$ on 
cosmological scales: $d \ln e^2 / dt \sim 10^{-10} \, {\rm yr}^{-1}$ (to be compared with 
the Oklo limit $\sim 5 \times 10^{-17} \, {\rm yr}^{-1}$ [see below], or with laboratory 
limits $\sim 10^{-14} \, {\rm yr}^{-1}$~\cite{PTM95,S01}).

It is generally assumed that this violent conflict with experimental tests of the 
equivalence principle is avoided because, after supersymmetry breaking, the 
dilaton\footnote{In the following, we use the word ``dilaton'' to denote the 
combination of the ten-dimensional dilaton and of various moduli which determines the 
values of the four-dimensional coupling constants.} might acquire a (large enough) 
mass: say $m_{\phi} \gtrsim 10^{-3} \, {\rm eV}$ so that observational deviations from 
Einstein's gravity are quenched on distances larger than a fraction of a millimeter. 
If that were the case, there would also be no possibility to predict any time 
variation of the coupling constants on cosmological scales. There exists, however, a 
mechanism which can naturally reconcile a {\it massless} dilaton with existing 
experimental data: this is the {\it decoupling mechanism} of Ref.~\cite{DP94} (see 
also~\cite{DN93}). In the following, we shall review a recent work~\cite{DPV1,DPV2} 
which has extended this mechanism in a manner which comes close to reconciling present 
experimental tests of the equivalence principle with the recent results of Webb et al.~\cite{W01} 
suggesting that the fine-structure constant $e^2$ has varied by $\sim 
10^{-5}$ between redshifts of order 1 and now. For recent reviews of the flurry of works
concerning the observational and theoretical aspects of the ``variation of constants'' 
see~\cite{U02}, and the recent book~\cite{VFC}. For a recent critical assessment of
the various methodologies for extracting a variation of the fine-structure
constant from astronomical data see~\cite{BSS03}.

\section{Decoupling mechanism and dilaton runaway}

The basic idea of Ref.~\cite{DP94} was to exploit the string-loop modifications of the 
(four dimensional) effective low-energy action (we use the signature $-+++$)
\begin{eqnarray}
\label{eq1.1}
S &= &\int d^4 x \sqrt{\widetilde{g}} \, \biggl( \frac{B_g (\phi)}{\alpha'} \, 
\widetilde R + \frac{B_{\phi} (\phi)}{\alpha'} \, \lbrack 2 \, \widetilde \Box \, \phi - 
(\widetilde{\nabla} \phi)^2 \rbrack \nonumber \\
&&- \frac{1}{4} \, B_F (\phi) \, \widetilde{F}^2 - \dots \biggl) \, ,
\end{eqnarray}
i.e. the $\phi$-dependence of the various coefficients $B_i (\phi)$, $i = g , \phi , 
F, \ldots$~, given in the weak-coupling region ($e^{\phi} \to 0$) by series of the 
form $B_i (\phi) = e^{-\phi} + c_0^{(i)} + c_1^{(i)} \, e^{\phi} + c_2^{(i)} \, 
e^{2\phi} + \cdots \, $, coming from the genus expansion of string theory. It was shown 
in~\cite{DP94} that, if there exists a special value $\phi_m$ of $\phi$ which extremizes 
all the (relevant) coupling functions $B_i^{-1} (\phi)$, the cosmological evolution of 
the graviton-dilaton-matter system naturally drives $\phi$ towards $\phi_m$ (which is 
a fixed point of the Einstein-dilaton-matter system). This provides a mechanism for 
fixing a massless dilaton at a value where it {\it decouples} from matter (``Least 
Coupling Principle''). Refs.~\cite{DPV1,DPV2} considered the case (recently suggested 
in~\cite{V01}) where the coupling functions, at least in the visible sector,
 have a smooth {\it finite} limit for infinite bare string coupling  $g_s \to \infty$. 
In this case, quite generically, we expect
\begin{equation}
\label{eq1.3}
B_i (\phi) = C_i + {\cal O} (e^{-\phi}) \, .
\end{equation}
Under this assumption, the coupling functions are all extremized at infinity, i.e. 
a fixed point of the cosmological evolution is $\phi_m = + \infty$. [See~\cite{GPV}
for an exploration of the late-time cosmology of models satisfying (\ref{eq1.3}).]
It was found that the ``decoupling'' of such a ``run-away'' dilaton   
has remarkable features: (i) the residual dilaton couplings at the
present epoch can be related to the amplitude of density fluctuations generated
during inflation, and (ii) these residual couplings, while being naturally
compatible with present experimental data, are predicted to be large enough to
be detectable by a modest improvement in the precision of equivalence principle
tests (non universality of the free fall, and, possibly, variation of ``constants'').
This result contrasts with the case of attraction towards a finite
value $\phi_m$ which leads to extremely small residual couplings~\cite{DV96}.

One assumes some primordial inflationary stage driven by the potential energy of an
inflaton field $\chi$. Working with the Einstein frame metric $g_{\mu \nu} = C_g^{-1} 
\, B_g (\phi) \, \widetilde{g}_{\mu \nu}$, and with the modified dilaton field 
$\varphi = \int d \phi [ (3/4) ( B'_g/B_g )^2 + B'_{\phi}/B_g + (1/2) \, 
B_{\phi}/B_{g} ]^{1/2}$, one considers an effective action of the form
\begin{eqnarray}
\label{eq2.4}
S &= &\int d^4 x \sqrt{g} \biggl[ \frac{\widetilde{m}_P^2}{4} \, R - 
\frac{\widetilde{m}_P^2}{2} \, (\nabla \varphi)^2 \nonumber \\
&&- \frac{\widetilde{m}_P^2}{2} \, F 
(\varphi) (\nabla \chi)^2 - \widetilde{m}_P^4 \, V (\chi , \varphi) \biggl] \, ,
\end{eqnarray}
where $\widetilde{m}_P^2 = 1/(4\pi G) = 4 C_g/ \alpha'$, and where the dilaton
dependence of the Einstein-frame action is related to its (generic) string-frame
dependence (\ref{eq1.1}) by $F(\varphi) = B_{\chi} (\phi) /$ $B_g (\phi) \, , \ V(\chi , 
\varphi) = C_g^{2} \, \widetilde{m}_P^{-4} \, B_g^{-2} (\phi) \, \widetilde V 
(\widetilde{\chi} , \phi)$.

Under the basic assumption (\ref{eq1.3}), $d\varphi / d \phi$ tends, in the 
strong-coupling limit $\phi \to +\infty$, to the constant $1/c$, with $c \equiv (2 C_g 
/ C_{\phi})^{1/2}$, so that the asymptotic behaviour of the bare string coupling is 
\begin{equation}
\label{eq2.6}
g_s^2 = e^{\phi} \simeq e^{c\varphi} \, .
\end{equation}
Let us consider for simplicity the case where $F(\varphi) = 1$ and $V (\chi , \varphi) 
= \lambda (\varphi) \, \chi^n/n$ with a dilaton-dependent inflaton coupling constant 
$\lambda (\varphi)$ of the form
\begin{equation}
\label{eq2.11}
\lambda (\varphi) = \lambda_{\infty} (1 + b_{\lambda} \, e^{-c\varphi}) \, ,
\end{equation}
where we assume that $b_{\lambda} > 0$, i.e that $\lambda (\varphi)$ reaches a {\it 
minimum} at strong-coupling, $\varphi \rightarrow + \infty$. It is shown in~\cite{DPV2} that this 
simple case is representative of rather general cases of 
$\varphi$-dependent inflationary potentials $V(\chi , \varphi)$.

During inflation ($ds^2 = -dt^2 + a^2 (t) \, \delta_{ij} \, dx^i \, dx^j$), it is 
easily seen that, while $\chi$ slowly rolls down towards $\chi \sim 1$, the dilaton 
$\varphi$ is monotically driven towards large values. The solution of the (classical) 
slow-roll evolution equations leads to  
\begin{equation}
\label{eq2.16} 
e^{c \varphi} + \frac{b_{\lambda} \, c^2}{2n} \, \chi^2 = {\rm const}. = e^{c 
\varphi_{\rm in}} + \frac{b_{\lambda} \, c^2}{2n} \, \chi_{\rm in}^2 \, .
\end{equation}
Using the result (\ref{eq2.16}), one can estimate the value $\varphi_{\rm end}$ of 
$\varphi$ at the end of inflation by inserting for the initial value $\chi_{\rm in}$ 
of the inflaton the value corresponding to the end of self-regenerating inflation~\cite{L90}. One 
remarks that the latter value can be related to the amplitude 
${\delta_H} \sim 5 \times 10^{-5}$ of density fluctuations, on the scale corresponding 
to our present horizon, generated by inflation, through $\chi_{\rm in} \simeq \, 5 
\sqrt n \, ({\delta_H} )^{ - 2/(n+2)}$. Finally, assuming $e^{c \varphi_{\rm in}} \ll 
e^{c \varphi_{\rm end}}$, one gets the estimate:
\begin{equation}
\label{eq2.23'}
e^{c \varphi_{\rm end}} \sim 12.5  c^2 \, b_{\lambda} \, \left(\delta_H\right)^{ 
-\frac{4}{n+2}} \, .
\end{equation}
A more general study~\cite{DPV2} of the run-away of the dilaton during inflation 
(including an estimate of the effect of quantum fluctuations) only modifies this 
result by a factor ${\cal O}(1)$. It is also found that the present value of the 
dilaton is well approximated by $\varphi_{\rm end}$.

\section{Deviations from general relativity induced by a runaway dilaton}

Eq.~(\ref{eq2.23'}) tells us that, within this scenario, the smallness of the present
matter couplings of the dilaton is quantitatively linked to the smallness of the
(horizon-scale) cosmological density fluctuations. To be more precise, and to 
study the compatibility with present experimental data, one needs to estimate the
crucial dimensionless quantity
\begin{equation}
\label{eq3.1}
\alpha_A (\varphi) \equiv \partial \ln m_A (\varphi) / \partial \, \varphi \, ,
\end{equation}
which measures the coupling of $\varphi$ to a massive particle of type $A$. 
The definition of $\alpha_A$ is such that, at the Newtonian approximation, the 
interaction potential between particle $A$ and particle $B$ is $-G_{AB} \, m_A \, m_B
/ r_{AB}$ where~\cite{DN93,DP94} $G_{AB} = G (1 + \alpha_A \, \alpha_B) $. Here, $G$
is the bare gravitational coupling constant entering the Einstein-frame action 
(\ref{eq2.4}), and the term $\alpha_A \, \alpha_B$ comes from the additional
attractive effect of dilaton exchange 
($ - G \, m_A \, m_B \alpha_A \, \alpha_B / r_{AB}$).
 
Let us first consider the (approximately) {\it composition-independent} deviations
from general relativity, i.e. those that do not essentially depend on violations of 
the equivalence principle. Most composition-independent gravitational experiments (in 
the solar system or in binary pulsars) consider the long-range interaction between 
objects whose masses are essentially baryonic (the Sun, planets, neutron stars). As 
argued in~\cite{TV88,DP94} the relevant coupling coefficient $\alpha_A$ is then 
approximately universal and given by the logarithmic derivative of the QCD confinement 
scale $\Lambda_{\rm QCD} (\varphi)$, because the mass of hadrons is essentially given 
by a pure number times $\Lambda_{\rm QCD} (\varphi)$. [We shall consider below the 
small, non-universal, corrections to $m_A (\varphi)$ and $\alpha_A (\varphi)$ 
linked to QED effects and quark masses.] Remembering from Eq.~(\ref{eq1.1}) the 
fact that, in the string frame (where there is a fixed cut-off linked to the string 
mass $\widetilde{M}_s \sim (\alpha')^{-1/2}$) the gauge coupling is dilaton-dependent 
($g_F^{-2} = B_F (\varphi)$), we see that (after conformal transformation) the 
Einstein-frame confinement scale has a dilaton-dependence of the form
\begin{equation}
\label{eq3.5}
\Lambda_{\rm QCD} (\varphi) \sim C_g^{1/2} \, B_g^{-1/2} (\varphi) \exp [- 8 \pi^2 \, 
b_3^{-1} \, B_F (\varphi)] \, \widetilde{M}_s \, ,
\end{equation}
where $b_3$ denotes the one-loop (rational) coefficient entering the Renormalization 
Group running of $g_F$. Here $B_F (\varphi)$ denotes the coupling to the ${\rm SU} 
(3)$ gauge fields. For simplicity, we shall assume that (modulo rational coefficients) 
all gauge fields couple (near the string cut off) to the same $B_F (\varphi)$. Such an 
assumption is natural in a stringy framework. Note that we differ here from the line 
of work of Jordan~\cite{J37} and Bekenstein~\cite{B82}, 
recently extended in~\cite{SBM02,OP02}, which
assumes that $\varphi$ couples {\it only} to the electromagnetic gauge field. The 
string-inspired assumption of coupling to all gauge fields yields the following 
approximately universal dilaton coupling to hadronic matter
\begin{equation}
\label{eq3.6}
\alpha_{\rm had} (\varphi) \simeq \left( \ln \left(\frac{\widetilde{M}_s}{\Lambda_{\rm
QCD}} \right) + \frac{1}{2} \right) \frac{\partial \ln B_F^{-1} (\varphi)}{\partial \,
\varphi} \, .
\end{equation}
Numerically, the coefficient in front of the R.H.S. of (\ref{eq3.6}) is of order 40. 
[For refinements on the estimate of this coefficient, i.e. on 
$\partial \ln\Lambda_{\rm QCD} / \partial \ln e^2$, see Ref.~\cite{L03}
and references therein.]
Consistently with the basic assumption (\ref{eq1.3}), one parametrizes 
the $\varphi$ dependence of the gauge coupling $g_F^2 = B_F^{-1}$ as
\begin{equation}
\label{eq3.7}
B_F^{-1} (\varphi) = B_F^{-1} (+ \infty) \, [1 - b_F \, e^{-c\varphi}] \, .
\end{equation}
We finally obtain 
\begin{equation}
\label{eq3.8}
\alpha_{\rm had} (\varphi) \simeq 40 \, b_F \, c \, e^{-c\varphi} \, .
\end{equation}
Inserting the estimate (\ref{eq2.23'}) of the value of $\varphi$ reached because of 
the cosmological evolution, one gets the estimate
\begin{equation}
\label{eq3.9}
\alpha_{\rm had} (\varphi_{\rm end}) \simeq 3.2 \, \frac{b_F}{b_{\lambda} \, c} \, 
\delta_H^{\frac{4}{n+2}} \, .
\end{equation}
It is plausible to expect that the quantity $c$ (which is a ratio) and the ratio $b_F 
/ b_{\lambda}$ are both of order unity. This then leads to the numerical estimate 
$\alpha_{\rm had}^2 \sim 10 \, \delta_H^{\frac{8}{n+2}}$, with $\delta_H \simeq 5 
\times 10^{-5}$. An interesting aspect of this result is that the expected present 
value of $\alpha_{\rm had}^2$ depends rather strongly on the value of the exponent $n$ 
(which entered the inflaton potential $V(\chi) \propto \chi^n$). In the case $n=2$ 
(i.e. $V(\chi) = \frac{1}{2} \, m_{\chi}^2 \, \chi^2$) we have $\alpha_{\rm had}^2 
\sim 2.5 \times 10^{-8}$, while if $n=4$ ($V (\chi) = \frac{1}{4} \, \lambda \, 
\chi^4$) we have $\alpha_{\rm had}^2 \sim 1.8 \times 10^{-5}$. Both estimates are 
compatible with present (composition-independent) experimental limits on deviations 
from Einstein's theory (in the solar system, and in binary pulsars).
For instance, the ``Eddington'' parameter $\gamma - 1 \simeq -2 \, \alpha_{\rm had}^2$ 
is compatible with the present best limits $\vert \gamma - 1 \vert \lesssim 2 \times 
10^{-4}$ coming from Very Long Baseline Interferometry measurements of the deflection 
of radio waves by the Sun~\cite{exp}.

Let us now consider situations where the non-universal couplings of the dilaton induce
(apparent) {\it violations of the equivalence principle}. This means considering the 
composition-dependence of the dilaton coupling $\alpha_A$, Eq.~(\ref{eq3.1}), i.e. the 
dependence of $\alpha_A$ on the type of matter we consider. Two test masses, made 
respectively of $A$- and $B$-type particles will fall in the gravitational field 
generated by an external mass $m_E$ with accelerations differing by
\begin{equation}
\label{eq3.14}
\left( \frac{\Delta a}{a} \right)_{AB} \equiv 2 \, \frac{a_A - a_B}{a_A + a_B} \simeq 
(\alpha_A - \alpha_B) \, \alpha_E \, .
\end{equation}
We have seen above that in lowest approximation $\alpha_A \simeq \alpha_{\rm had}$ 
does not depend on the composition of $A$. We need, however, now to retain the small 
composition-dependent effects to $\alpha_A$ linked to the $\varphi$-dependence of QED 
and quark contributions to $m_A$. This has been investigated in~\cite{DP94} (see also~\cite{DD03} 
for a study of the $\varphi$-dependence of the quark contributions to nuclear binding energies) 
with the result that $\alpha_A - \alpha_{\rm had}$ depends linearly on the baryon number $B 
\equiv N + Z$, the neutron excess $D \equiv N-Z$, and the quantity $E \equiv Z (Z-1) / 
(N+Z)^{1/3}$ linked to nuclear Coulomb effects. [The standard ``adiabatic'' way of estimating the 
$\varphi$-dependence of $m_A$ has been questioned in~\cite{B02}. We show in Section~5 below that 
the claims of~\cite{B02} are both unjustified and phenomenologically excluded.] Under the 
assumption that the latter 
dependence is dominant, and using the average estimate $\Delta (E/M) \simeq 2.6$ 
(applicable to mass pairs such as (Beryllium, Copper) or (Platinum, Titanium)), one 
finds that the violation of the universality of free fall is approximately given by
\begin{equation}
\label{eq3.17}
\left( \frac{\Delta a}{a} \right) \simeq 5.2 \times 10^{-5} \, \alpha_{\rm had}^2 
\simeq 5.2 \times 10^{-4} \left( \frac{b_F}{b_{\lambda} \, c} \right)^2 \, 
\delta_H^{\frac{8}{n+2}} \, .
\end{equation}
This result is one of the main predictions of the present model. If one inserts the 
observed density fluctuation $\delta_H \sim 5 \times 10^{-5}$, one obtains a level 
of violation of the universality of free fall (UFF) due to a run-away dilaton which is
$\Delta a/a \simeq 1.3 ( b_F/(b_{\lambda} \, c) )^2 \times 10^{-12}$ for  $n=2$ (i.e.
for the simplest chaotic inflationary potential $V (\chi) = \frac{1}{2} \, m_{\chi}^2 
(\phi) \, \chi^2$), and $\Delta a/a \simeq 0.98 ( b_F/(b_{\lambda} \, c) )^2 \times 
10^{-9}$ for $n=4$ (i.e. for $V (\chi) = \frac{1}{4} \, \lambda (\phi) \, \chi^4$).
The former case is naturally compatible with current tests (at the $ \sim 10^{-12}$ 
level~\cite{Su94}) of the UFF. Values $ n \geq 4$ of the exponent are somewhat disfavoured
(within this scenario) because they would require that the (unknown) dimensionless 
combination of parameters 
$(b_F/(b_{\lambda} \, c) )^2$ be significantly smaller than one. It is
interesting to remark that the recent WMAP observational results also
disfavour values $ n \geq 4$ of the inflationary-potential exponent~\cite{WMAP}.

\section{Cosmological variation of ``constants''}

Let us also consider another possible deviation from general relativity and the 
standard model: a possible time variation of the coupling constants, most notably of the
fine structure constant $e^2/\hbar c$ on which the strongest limits are available.
Consistently with our previous assumptions we expect $e^2 \propto B_F^{-1} (\varphi)$ 
so that, from (\ref{eq3.7}), $e^2 (\varphi) = e^2 (+ \infty) \, [1 - b_F \, e^{- c
\varphi} ]$. The logarithmic variation of $e^2$ (introducing the  derivative
$\varphi' = d \varphi / d p$ with respect to the ``e-fold'' parameter $dp = H \, dt = 
d a/ a $) is thus given by
\begin{equation}
\label{eq3.21bis}
\frac{d \ln e^2}{H \, dt}  \simeq b_F \, c \, e^{-c \varphi} \, 
\varphi' \, \simeq \frac{1}{40} \alpha_{\rm had} \varphi' \, .
\end{equation}
The value of $\varphi'$ depends on the coupling of the dilaton to the two currently 
dominating energy forms in the universe: dark matter (coupling $\alpha_m (\varphi)$), 
and vacuum energy (coupling $\alpha_V =  \frac{1}{4} \, \partial \ln V(\varphi) / 
\partial \, \varphi$). In the slow-roll approximation, the cosmological evolution of 
$\varphi$ is given by
\begin{equation}
\label{eq2.32bis}
(\Omega_m + 2 \Omega_V ) \varphi' = - \Omega_m \alpha_m - 4 \Omega_V \alpha_V \, ,
\end{equation}
where $\Omega_m $ and $\Omega_V$ are, respectively, the dark-matter- and the 
vacuum-fraction of critical energy density ($\rho_c \equiv (3/2) {\widetilde{m}_P^2} 
H^2$). The precise value of $\varphi'$ is model-dependent and can vary (depending 
upon the assumptions one makes) from an exponentially small value ($ \varphi' \sim
e^{- c \varphi}$) to a value of order unity. In models where either the dilaton is 
more strongly coupled to dark matter than to ordinary matter~\cite{DGG}, or/and plays 
the role of quintessence (as suggested in~\cite{GPV}), $\varphi'$ can be of order 
unity. Assuming just spatial flatness and saturation of the ``energy budget'' by
non-relativistic matter and dilatonic quintessence, one can relate the value of 
$\varphi' = d \varphi / (H dt)$ to $\Omega_m $ and to the deceleration parameter $q 
\equiv - \ddot{a}a/\dot{a}^2$:
\begin{equation}
\label{q2}
\varphi'^2 = 1 + q - \frac{3}{2} \Omega_m.
\end{equation}
The supernovae Ia data~\cite{SNI} give a strict upper bound on the present value 
$q_0$: $q_0<0$. A generous lower bound on the present value of $\Omega_m$ is 
$\Omega_{m 0} > 0.2$~\cite{omegam}. Inserting these two constraints in Eq.(\ref{q2}) 
finally yields the safe upper bound on the current value of $\varphi'$
\begin{equation}
\label{q3}
{\varphi'_0}^2 < 0.7 \, , \; {\rm i.e.} \; \vert \varphi'_0 \vert < 0.84 \; .
\end{equation}
On the other hand, Eq.~(\ref{eq3.17}) yields the link
\begin{equation}
\label{eqn2}
\alpha_{\rm had} \simeq \pm \, 1.4 \times 10^{-4} \, \sqrt{10^{12} \, \frac{\Delta a}{a}} 
\, .
\end{equation}
Inserting this result in Eq.~(\ref{eq3.21bis}) yields
\begin{equation}
\label{eqn3}
\frac{d \ln e^2}{H dt} \simeq \pm \, 3.5 \times 10^{-6} \, \sqrt{10^{12} \, \frac{\Delta 
a}{a}} \, \varphi' \, ,
\end{equation}
which yields, upon integration over $p = \ln a + {\rm cst} = - \ln (1+z) + {\rm cst}$,
\begin{equation}
\label{eqn4}
\frac{\Delta e^2}{e^2} \equiv \frac{e^2 (z) - e^2 (0)}{e^2 (0)} \simeq \mp \, 3.5 \times 
10^{-6} \, \sqrt{10^{12} \, \frac{\Delta a}{a}} \, \langle \varphi' \rangle_z \, \ln 
(1+z) \, ,
\end{equation}
where $\langle \varphi' \rangle_z \equiv (\varphi (p) - \varphi (p_0)) / (p-p_0)$ 
denotes the average value of $\varphi'$ between now and redshift $z$. If we insert in 
Eq.~(\ref{eqn3}) the limit $\Delta a / a \lesssim 10^{-12}$, coming from present 
experimental tests of the universality of free fall (UFF)~\cite{Su94}, as well as the 
cosmological constraint (\ref{q3}) on the present value of $\varphi'$, we find that 
the present variation of the fine-structure constant is constrained to be $\vert d \ln 
e^2 / H dt \vert \lesssim 3 \times 10^{-6}$, i.e. $\vert d \ln e^2 / dt \vert \lesssim 
2 \times 10^{-16} \, {\rm yr}^{-1}$. Such a level of variation is comparable to the planned 
sensitivity of currently developed cold-atom clocks~\cite{S01}.

However, there are stronger constraints coming from geochemical data. Let us first 
recall that a {\it secure} limit on the time variation of $e^2$ coming from the Oklo 
phenomenon is $\vert \Delta e^2 / e^2 \vert \lesssim 10^{-7}$ between now and $\sim 2 
\, {\rm Gyr}$ ago, i.e. $\vert d \ln e^2 / dt \vert \lesssim 5 \times 10^{-17} \, {\rm 
yr}^{-1}$. This 
limit was obtained in~\cite{DD96} under two very conservative assumptions: 1. A 
conservative interpretation of Oklo data making minimal assumptions about the 
temperature of the reaction zone, and about the possible amplitude of variation of the 
resonance energy $E_r$ of the relevant excited state of Samarium 150, and 2. An 
analysis of the $e^2$-variation of the latter resonance energy $E_r$ taking into 
account only the (rather well-known) Coulomb effects. We note that Ref.~\cite{Fujii00} 
derived a stronger limit on the variation of $e^2$ by replacing assumption 1. above by 
the non-conservative assumption that $E_r$ has stayed close to its present value. We 
note also that Ref.~\cite{OPQCCV} derived a stronger limit on the variation of $e^2$ 
by relaxing assumption 2. and by trying to estimate the indirect $e^2$-dependence of 
$E_r$ coming from light quark contributions to the nuclear binding energy.

It is important to note that Ref.~\cite{OPQCCV} also derived the strong limit $\vert 
\Delta e^2 / e^2 \vert \lesssim 3 \times 10^{-7}$ between now and $4.6 \, {\rm Gyr}$ ago. This 
limit was derived from the Rhenium/Osmium ratio in $4.6 \, {\rm Gyr}$ old iron-rich meteorites 
by making quite conservative assumptions. In particular, we note that, similarly to 
the Oklo assumption 2. above, this limit uses only the Coulomb effects in the 
$\beta$-decay rate of Rhenium 187. [Ref.~\cite{OPQCCV} quotes also stronger limits 
obtained by trying to estimate the indirect $e^2$-dependence of the latter 
$\beta$-decay rate.]

Using Eq.~(\ref{eqn4}), the conservative ``Oklo'' and ``Rhenium'' limits on the 
variation of $e^2$, corresponding to redshifts $z_{\rm Oklo} \simeq 0.15$ and $z_{\rm 
Re} \simeq 0.45$, can be converted into constraints on the product $r \, \langle 
\varphi' \rangle_z$ where $r \equiv \sqrt{10^{12} \, \Delta a / a}$. In round numbers, 
one finds that these two constraints read
\begin{equation}
\label{eqn5}
\sqrt{10^{12} \, \frac{\Delta a}{a}} \, \vert \langle \varphi' \rangle_{z \simeq 0.15}
\vert \lesssim 0.2 \, ; \ \sqrt{10^{12} \, \frac{\Delta a}{a}} \, \vert \langle 
\varphi' \rangle_{z \simeq 0.45} \vert \lesssim 0.2 \, .
\end{equation}

Rigourously speaking, while the first (Oklo) constraint above does involve
only a difference between redshift $z \simeq 0.15$ and redshift $ z=0$ (
because the Oklo phenomenon took place 2 billion years ago), the second (Rhenium)
constraint should involve some (model-dependent) average over redshifts
$ 0 \leq z \leq 0.45$. However, for simplicity (and in view of the approximate nature
of the Rhenium constraint), we shall work with a Rhenium constraint also
expressed as a difference between $z=0.45$ and $z=0$.

As seen on Eq.~(\ref{eq2.32bis}), the cosmological evolution of $\varphi$ is driven by 
two quantities: the coupling $\alpha_m$ of $\varphi$ to dark matter, and its coupling 
$\alpha_V$ to vacuum energy. If one had only the Oklo constraint to cope with, one 
could fine-tune the {\it ratio} $\alpha_m / \alpha_V$ to satisfy the first constraint 
(\ref{eqn5}) without constraining the overall magnitudes of $\alpha_m$ and $\alpha_V$. 
This is what was done in Ref.~\cite{OP02} (within the different context of 
Jordan-Bekenstein-like models) to exhibit models satisfying the Oklo and UFF constraints and
allowing for a variation of $e^2$ around $z \sim 1$ driven by a large enough 
$\alpha_m$ (\`a la~\cite{DGG90}) to explain the observational results of Webb et al.~\cite{W01}. 
[Note that the implementation of the same idea within the context of 
dilaton-like models~\cite{DPV1,DPV2} leads to a maximal possible variation of $e^2$ 
which falls short, by a factor $\sim 4$, of the level needed to explain the results 
of~\cite{W01}.] However, the point we wish to emphasize here is that the recently 
obtained ``Rhenium constraint''~\cite{OPQCCV}, i.e. the second inequality 
(\ref{eqn5}), makes it impossible to concoct a fine-tuned ratio $\alpha_m / \alpha_V$ 
so as to satisfy the two geochemical constraints (\ref{eqn5}), which correspond to 
quite different redshifts and therefore to significantly different relative weights 
$\Omega_m \propto (1+z)^3$ and $\Omega_V \propto (1+z)^0$. If we were to consider more 
complicated models, in particular models where the ``kinetic term'' $\propto \, 
\varphi''$ must be included in Eq.~(\ref{eq2.32bis}), and allows for an 
oscillatory behaviour (as in local-attractor models~\cite{DP94}), it might be possible 
to fine-tune more parameters (like the phase of oscillation of $\varphi$) so as to 
satisfy the two constraints (\ref{eqn5}). An example of a model with more parameters
(namely, the mass of the scalar field) which can be tuned to minimize the variation of
$e^2$ over redshifts $ 0 \leq z \leq 0.45$ while allowing a $ \sim 10^{-5}$
variation for higher redshifts has been recently given~\cite{G03}.
However, such a heavy fine-tuning becomes
very unnatural. The most natural conclusion is that both $\alpha_m$ and $\alpha_V$ 
must be small enough, so that the quantity $\sqrt{10^{12} \, \Delta a/a} \, \vert 
\langle \varphi' \rangle_z \vert$ is smaller than $0.2$ for all redshifts where the 
cosmologically dominant energy forms are dark matter and/or dark energy. 
Eq.~(\ref{eqn4}) then leads to the constraint
\begin{equation}
\label{eqn6}
\frac{\vert e^2 (z) - e^2 (0) \vert}{e^2 (0)} \lesssim 0.7 \times 10^{-6} \ln (1+z) \, 
.
\end{equation}
Note that this constraint is about ten times smaller than the claim of~\cite{W01}.

\section{On a claim of Bekenstein}

Recently, Bekenstein~\cite{B02} has claimed that, within the context of the Jordan 
model~\cite{J37} (revived in~\cite{B82}) where $\varphi$ couples only to the electromagnetic 
gauge 
field, the variability of the fine-structure constant $e^2$ implies no detectable violations of 
the weak equivalence principle. This claim is based on a two-step argument.

First, Ref.~\cite{B02} claims that the theoretical consistency of the model necessarily implies a 
very particular dependence of particle masses on the scalar field. Namely, the mass of 
electrically neutral elementary particles must be independent of the scalar field, while the mass 
of electrically charged ones must depend on $\psi$ in the following way:
\begin{equation}
\label{eqnn1}
m_A (\psi) = m_A^0 + \frac{e_A^0}{\kappa} \, [{\rm arcsec} \, e^{\psi} - \sqrt{e^{2\psi} -1}] \, 
,
\end{equation}
so that
\begin{equation}
\label{eqnn2}
\frac{\partial \, m_A (\psi)}{\partial \, \psi} = - \frac{e_A^0}{\kappa} \ \sqrt{e^{2\psi} - 1} 
\, .
\end{equation}
We are here using the notation of~\cite{B02}. In particular, the scalar field is denoted $\psi$ 
and is normalized so that its (Einstein-frame) kinetic term is $- (8 \pi \, \kappa^2)^{-1} \, 
(\nabla \psi)^2$ and the coupling to the electromagnetic field is $-(16\pi)^{-1} \, e^{-2\psi} 
F_{\mu\nu}^2$. The link with the (Einstein-frame) notation used above is $\psi = \kappa \,
\varphi / \sqrt G$ and $B_F (\varphi) = \exp (-2\psi) = \exp (-2 \kappa \, \varphi / \sqrt G)$. 
The scalar coupling constant $\kappa^2$ has the same dimension as $G$ and is supposed not to be 
very different from $G$.

Second, Ref.~\cite{B02} claims that the specific scalar-dependence (\ref{eqnn1}), (\ref{eqnn2}) 
entails a cancellation between Coulomb and scalar forces which ensure the validity of the 
equivalence principle.

We think that both claims are untenable. Concerning the first claim, namely the {\it necessity}
of the specific scalar dependence (\ref{eqnn1}), the reasoning of~\cite{B02} is based on the 
identification of the coefficients of (three-dimensional) delta-function terms in some field
equations. However, this identification is physically and mathematically unjustified because 
these equations contain, besides the $\delta^3 ({\bf x})$ source terms, some terms proportional 
to the Coulomb energy ${\bf E}^2 (x)$ of the considered classical charged point particle. Such a 
distributed source term is non locally integrable ($\int d^3 x \, r^{-4} = \infty$) and is even 
(classically) more divergent than the critical power-law $(r^{-3})$ whose presence already 
signals (in renormalization theory) the possibility to add an arbitrary multiple of $\delta^3 
({\bf x})$. In physical terms, the infinite contribution, proportional to the Coulomb 
self-energy, makes it meaningless to keep track of the ``bare'' $\delta^3 ({\bf x})$ source terms 
in the scalar field equation. A correct treatment of the infinite Coulomb self-energy calls 
either for the explicit consideration of a (Poincar\'e-like) finite classical model of an 
extended electron, or for the application of (classical or quantum) renormalization theory. In 
both cases, the ground for the identification of bare $\delta^3 ({\bf x})$ source terms will 
disappear, and I expect that the standard coupling of $\psi$ to the (finite) renormalized Coulomb 
self-energy will remain. Such a coupling is well-known to entail violations of the equivalence 
principle.

Independently of the above argument concerning the lack of necessity of the scalar dependence 
(\ref{eqnn1}), let us now show that Eq.~(\ref{eqnn1}) is already phenomenologically excluded.

Let us first recall (in a different language) the argument of~\cite{B02} concerning the 
cancellation taking place in two-body interactions. Let $- K_{AB}^s / r_{AB}$ denote the 
interaction energy between two particles {\it at rest} due to the exchange of a spin-$s$ field. 
For scalar $(s=0)$ exchange we have $K_{AB}^0 = + \, G \, m_A \, m_B \, \alpha_A \,\alpha_B$
(see above), i.e.
\begin{equation}
\label{eqnn3}
K_{AB}^0 = + \, G \, \frac{\partial \, m_A}{\partial \, \varphi} \, \frac{\partial \,
m_B}{\partial \, \varphi} = + \, \kappa^2 \, \frac{\partial \, m_A}{\partial \, \psi} \, 
\frac{\partial \, m_B}{\partial \, \psi} \, .
\end{equation}
The insertion of (\ref{eqnn2}) into (\ref{eqnn3}) then yields
\begin{equation}
\label{eqnn4}
K_{AB}^0 = e_A^0 \, e_B^0 (e^{2 \psi_{\infty}} - 1) \, ,
\end{equation}
in which $\psi_{\infty}$ denotes the ``VEV'' of $\psi$, i.e. the value it takes far from all 
localized sources. In addition to the scalar interaction (\ref{eqnn4}) one needs to consider the 
Coulomb interaction ($s=1$; with a vacuum permittivity $4 \pi \, \epsilon_0 = e^{-2 
\psi_{\infty}}$)
\begin{equation}
\label{eqnn5}
K_{AB}^1 = - e_A^0 \, e_B^0 \, e^{2 \psi_{\infty}} \, ,
\end{equation}
and the gravitational one ($s=2$)
\begin{equation}
\label{eqnn6}
K_{AB}^2 = + \, G \, m_A (\psi_{\infty}) \, m_B (\psi_{\infty}) \, .
\end{equation}
The cancellation emphasized in~\cite{B02} is the cancellation of the $\psi_{\infty}$-dependence 
in the sum of (\ref{eqnn4}) and (\ref{eqnn5}),
\begin{equation}
\label{eqnn7}
K_{AB}^0 + K_{AB}^1 = - e_A^0 \, e_B^0 \, ,
\end{equation}
which gives back the standard ($\psi_{\infty}$-independent) Coulomb interaction between two 
(constant) charges $e_A^0$.

The second main objection that we wish to raise here is that the result (\ref{eqnn4}) is valid 
(when granting the first assumption (\ref{eqnn1})) only for ``elementary'' charged particles {\it 
at rest}, and is strongly modified when considering ``composite'' particles, whose structure 
comprise {\it relativistically moving} charged particles. For definiteness, one can have in mind 
an atom, viewed as a (classical) collection of $Z$ protons and $Z$ electrons. In the classical 
framework assumed in~\cite{B02} the part of the source of the scalar field which is localized
on elementary particles is
\begin{eqnarray}
\label{eqnn8}
\sqrt g \, \Box_g \, \psi &= &4 \pi \, \kappa^2 \sum_A \frac{\partial \, m_A}{\partial \psi} \, 
\frac{ds_A}{dt} \, \delta^3 ({\bf x} - {\bf z}_A (t)) + \cdots \nonumber \\
&= &-4\pi \, \kappa \, \sqrt{e^{2\psi} - 1} \sum_A e_A^0 \, \frac{ds_A}{dt} \, \delta^3 ({\bf x} 
- {\bf z}_A (t)) + \cdots \, ,
\end{eqnarray}
where $ds_A = (-g_{\mu\nu} (z_A) \, dz_A^{\mu} \, dz_A^{\nu})^{1/2}$ is the proper time along the 
worldline of the $A^{\rm th}$ elementary charge $e_A^0$ present in the considered composite 
object. Neglecting general relativistic effects and focussing on special relativistic ones 
$(g_{\mu\nu} = \eta_{\mu\nu})$, Eq.~(\ref{eqnn8}) shows that the coupling of a composite object
(i.e. a localized collection of elementary particles)
to the scalar field is not described by the total electric charge $\Sigma \, e_A^0$ present in 
the object, but instead by an ``effective charge'' $\bar e$ given by
\begin{equation}
\label{eqnn9}
\bar e = \sum_A \, e_A^0 \, \frac{ds_A}{dt} = \sum_A e_A^0 \, \sqrt{1 - {\bf v}_A^2} \, .
\end{equation}
In view of Eq.~(\ref{eqnn8}) it is this effective charge which enters in the scalar-matter 
coupling as well as in the scalar field energy. Therefore the effective scalar interaction 
between two composite objects, say $\bar A$ and $\bar B$, will read (when the composite objects 
are globally at rest with respect to each other)
\begin{equation}
\label{eqnn10}
K_{\bar A \bar B}^0 = \bar{e}_{\bar A} \, \bar{e}_{\bar B} (e^{2\psi_{\infty}} - 1) \, .
\end{equation}
If we consider, for definiteness, an atom, with internal velocities for the electrons 
proportional to $Z$ times the fine-structure constant $\alpha_{\rm em} \equiv e^2 / \hbar$ ($v_n 
= Z \, \alpha_{\rm em} / n$ for the $n^{\rm th}$ classical Bohr orbit), we see from 
(\ref{eqnn9}), that an electrically neutral atom will have a non vanishing effective (scalar) 
charge $\bar e \simeq \displaystyle{\sum_A} \, e_A^0 (1 - {\bf v}_A^2 / 2) \simeq - 
\displaystyle{\sum_A} \, e_A^0 \, {\bf v}_A^2 / 2$, of order $\bar e \sim Z^2 \, \alpha_{\rm 
em}^2 \vert e \vert$ where $-\vert e \vert$ is the charge of the electron. From (\ref{eqnn10}) we 
then deduce that the specific scalar dependence (\ref{eqnn1}) implies a scalar attraction between 
atoms of order $K_{\bar A \bar B}^0 \sim + Z_{\bar A}^2 \, Z_{\bar B}^2 \, \alpha_{\rm em}^4 \, 
e^2 (e^{2 \psi_{\infty}} - 1)$. Such a residual interaction between electrically neutral 
composite objects differs from what would be the unscreened electric interaction $\sim Z_{\bar A} 
\, Z_{\bar B} \, e^2$ by the factor $Z_{\bar A} \, Z_{\bar B} \, \alpha_{\rm em}^4 (e^{2 
\psi_{\infty}} - 1)$. It is useful to express such an additional electric-strength attraction 
between atoms in terms of the usual gravitational interaction. Using $e^2 \sim 10^{36} \, Gm_p$, 
where $m_p$ is the proton mass, and $\alpha_{\rm em}^4 \sim 10^{-8}$, we see that the assumption 
(\ref{eqnn1}) implies the existence of a {\it scalar attraction between atoms which is} $\sim 10^{28}
(e^{2\psi_{\infty}} - 1)$ {\it stronger than gravity}. This interaction will also violate the
equivalence principle. Even if we forget about equivalence-principle violations, the 
composition-independent tests of relativistic gravity exclude the existence of such an 
interaction, except if $e^{2 \psi_{\infty}} - 1 \lesssim 10^{-32}$. Such a level is about 27 
orders of magnitude smaller than the level $e^{2 \psi_{\infty}} - 1 \sim 10^{-5}$ considered by 
Bekenstein in connection with the results of Webb et al. Actually, the situation is worse if one 
takes into account the known composite nature of nuclei (made of nucleons with squared velocities
of a few percent), and even of protons and neutrons (made of relativistically moving quarks).

To conclude, the well-known fact that, contrary to the electric charge, the scalar charge of a 
relativistically moving particle is multiplied by the Lorentz factor $ds/dt = \sqrt{1 - {\bf 
v}^2}$ shows by itself the phenomenological impossibility of the scalar dependence (\ref{eqnn1}). 
In addition, the lack of ``stability'' of (\ref{eqnn1}) under the possible compositeness of 
charged ``particles'' is related to our first objection above pointing out the inconsistency of 
the ``derivation'' of (\ref{eqnn1}) based on the consideration of bare $\delta$-function source 
terms in presence of stronger singularities.

\section{Conclusions}

A first conclusion is that the results of Webb et al.~\cite{W01} cannot be naturally 
explained within any model where the time variation of the fine-structure constant 
$e^2$ is driven by the spacetime variation of a very light scalar field $\varphi$. On 
the other hand, the recently explored~\cite{DPV1,DPV2} class of dilaton-like models 
with an attractor ``at infinity'' in field space naturally predicts the existence of 
small, but not unmeasurably small, violations of the equivalence principle. In the 
case where the dilaton $\varphi$ is more significantly coupled to dark matter and/or 
dark energy than to ordinary matter, dilaton-like models can lead to a cosmological 
variation of $e^2$ as large as Eq.~(\ref{eqn6}). [We recall that this upper bound 
takes into account the two stringent geochemical bounds on $\Delta e^2 / e^2$ coming 
from the conservative interpretations of Oklo data~\cite{DD96} and of Rhenium decay 
data~\cite{OPQCCV}.] A time variation of the order of the upper limit in Eq.~(\ref{eqn6}) 
might be observable through the comparison of high-accuracy cold-atom clocks. It might 
also be observable in astronomical spectral data (if one understands how to explain 
and subtract the systematic effects leading to the current apparent variation of $e^2$ 
at the level $\Delta e^2 / e^2 \simeq -0.7 \times 10^{-5}$~\cite{W01}).

Finally, an important conclusion of all theoretical models where the time variation of 
$e^2$ is linked to the spacetime variation of a light scalar field $\varphi$ (be it 
the dilaton of string theory~\cite{DP94,DPV1,DPV2} or a field constrained to couple 
only to electromagnetism~\cite{J37,B82,SBM02,OP02}) is that a {\it necessary} condition
for having a fractional variation of $e^2$ larger than $10^{-6}$ on cosmological time 
scales is to have a violation of the universality of free fall (UFF) larger than about 
$10^{-13}$ (see Eq.~(\ref{eqn3}) in which $\vert \varphi' \vert$ is certainly 
constrained by cosmological data to be smaller than 1, as discussed in~\cite{DPV2}). 
Note that a measurably large violation of the UFF is a necessary, but by no means 
sufficient, condition for having a measurably large cosmological variation of $e^2$. 
Indeed, in Eq.~(\ref{eqn3}) or Eq.~(\ref{eqn4}) the value of $\varphi' \equiv d 
\varphi / (Hdt)$ depends on the strength of the coupling of $\varphi$ to the dominant 
forms of energy in the universe: $\alpha_m$ and $\alpha_V$. These quantities could 
well be comparable to the strength of the coupling of $\varphi$ to hadronic matter, 
$\alpha_{\rm had}$, which is contrained to be small by UFF experiments (see 
Eq.~(\ref{eqn2})). This shows that the best experimental probe of an eventual 
``variation of constants'' is to probe their {\it spatial} variation through 
high-precision tests of the UFF, rather than their (cosmological) time variation (see 
also~\cite{D00} for a detailed discussion of clock experiments). This gives additional 
motivation for improved tests of the UFF, such as the
Centre National d'Etudes Spatiales (CNES) mission MICROSCOPE~\cite{Touboul} (to fly in 
2005; planned sensitivity: $\Delta a / a \sim 10^{-15}$), and  the National 
Aeronautics and Space Agency (NASA) and European Space Agency (ESA) mission STEP 
(Satellite Test of the Equivalence Principle; planned sensitivity: $\Delta a / a \sim 
10^{-18}$)~\cite{worden}.

\section{Acknowledgements}
It is a pleasure to thank John Bahcall and John Donoghue for
informative discussions.

\end{document}